# Synthesis and superconductivity of CeNi$_{0.8}$Bi$_2$: New entrant in superconductivity kitchen?

Anuj Kumar, Shiva Kumar, Rajveer Jha and V.P.S. Awana[*]

Quatum Phenomena and Applications Division, National Physical Laboratory, Dr. K.S. Krishnan Marg, New Delhi 110012, India

CeNi$_{0.8}$Bi$_2$ is very recent entrant in superconductivity kitchen [1], and the same seeks reproduction from other groups. We report synthesis, structural details and superconductivity in CeNi$_{0.8}$Bi$_2$ compound. The CeNi$_{0.8}$Bi$_2$ compound is synthesized by solid state reaction route via vacuum encapsulation of high purity mixed Ce, Ni, Bi and subsequent heating of the same at 500 $^0$C for 10 hours followed by annealing at 750 $^0$C for 20 hours. As synthesized compounds are dark grey in color and do crystallize in tetragonal structure with space group *P*4/*nmm*. The lattice parameters are $a = 3.61$ Å and $c = 10.20$ Å. AC susceptibility measurements showed that synthesized CeNi$_{0.8}$Bi$_2$ compound is weakly superconducting below 4.2K. We believe our article will regenerate fresh activity to look for bulk superconductivity in CeNi$_{0.8}$Bi$_2$.

[*] Corresponding Author
Dr. V.P.S. Awana
Fax No. 0091-11-45609310: Phone No. 0091-11-45609357
e-mail-awana@mail.nplindia.ernet.in:
Research Home page - www.freewebs.com/vpsawana/

Undoubtedly, a hot pursuit is on for the search of new superconductors, after the discovery of superconductivity at 26 K in LaO$_{1-x}$F$_x$FeAs [2]. The substitution of La by other rare earth elements such as Ce, Pr, Sm, Nd, Gd and Tb led to a family of 1111 phase superconductors with $T_c$ of up to 56 K [2-4]. Soon after another interesting FeAs based family with slightly different structure namely 122 ($AFe_2As_2$: A alkaline metal) was invented [5,6] and superconductivity of up to 30 K could be achieved. This was further supplemented with observation of Superconductivity in $FeSe_{1-x}$ system at ~8 K in its tetragonal form in the absence of doping [7]. This compound is nick named as 11. Superconductivity in $FeSe_{1-x}$ system gets significantly affected by applied pressure ($dT_c/dP$ of around 9K/GPa) or chalcogenide substitutions (Se/Te) [8, 9]. Interestingly, FeTe is no longer superconducting. As far as the effect of magnetic field is concerned on the superconductivity of 1111, 122 or 11 systems, the 11 is found to be most robust [2-8]. Due to the recent [2] renewed interest in superconducting materials more superconductors keep on added [10]. Very recently, a similar structure (ZrCuSiAs-type) Ni containing compound ($CeNi_{0.8}Bi_2$) is reported [1] from the consortium of the oxy-pnictide inventors [2] group. Despite the gold rush for new superconducting materials, $CeNi_{0.8}Bi_2$ yet seeks its reproduction by other independent groups. May it be, that because of the relatively low (~4 K) superconducting transition temperature ($T_c$); in $CeNi_{0.8}Bi_2$ not many are interested. However, its structural similarity to high $T_c$ oxy-pnictides along with its rich crystal chemistry having two different types of Bi ions: Bi(1) forming $NiBi_4$ and Bi(2) as a Bi-square net, certainly warrant attention. In current communication, we report the synthesis and magnetization of the very recently discovered [1] $CeNi_{0.8}Bi_2$ superconductor. The synthesized $CeNi_{0.8}Bi_2$ is crystallized in tetragonal structure with space group $I4/mmm$ and is weakly superconducting below 4.2K.

$CeNi_{0.8}Bi_2$ compound is synthesized by solid state reaction route via vacuum encapsulation. High purity Ce, Ni, Bi are weighed in right stoichiometric ratio and ground thoroughly. It is to be noted that weighing and grinding was done in the glove box under high purity argon atmosphere. The powder was subsequently palletized and vacuum-sealed ($10^{-4}$ Torr) in a quartz tube. Sealed quartz ampoule was placed in box furnace and heat treated at 500 $^0$C for 10 hours followed by annealing at 750 $^0$C for 20 hours. Finally furnace was allowed to cool naturally. The X-ray diffraction pattern of the compound was taken on Rigaku diffractometer. The magnetization (*AC*) measurements were carried out using Quantum Design *PPMS* (Physical Property Measurement System).

As synthesized sample is grey color. One of the first samples caught fire due to its exposure to air. Later for freshly synthesized new samples, we opened the sealed quartz tube in Ar filled glove box with oxygen level below 1ppm, and took the sample immediately for powder X-ray diffraction (XRD). The room temperature X-ray diffraction (XRD) pattern for synthesized $CeNi_{0.8}Bi_2$ compound is shown in Figure 1. The compound is crystallized in tetragonal structure with space group *P*4/*nmm*. The lattice parameters are $a = 3.61$Å and $c = 10.20$ Å. It can be concluded from XRD results that the synthesized $CeNi_{0.8}Bi_2$ is nearly single phase but with some unidentified impurities.

AC susceptibility versus temperature $\chi(T)$ behavior of the $CeNi_{0.8}Bi_2$ sample is exhibited in Figure 2. AC susceptibility is done at 333Hz and 1Oe drive field. Worth mentioning is the fact, that powder of the sample being taken is filled in measuring capsule in Glove box itself in presence of high purity Ar and oxygen, humidity level below 1ppm. This is necessary to keep sample away from exposure to air/oxygen. We have successfully synthesized near single phase $CeNi_{0.8}Bi_2$ compound, which is weakly superconducting below 4.2K. Our results to some extent could be taken as the reproduction of the much warranted results of ref. 1. Further our short communication will stimulate the superconductivity researcher's community to pursue the newest superconducting materials kitchen entrant $CeNi_{0.8}Bi_2$.

Authors from NPL would like to thank Prof. R.C. Budhani (DNPL) for his constant encouragement. Anuj Kumar and Shiva Kumar are thankful to CSIR for providing the financial support during their research.

**Figure Captions**

Figure 1(a): Room temperature X-ray diffraction patterns of $CeNi_{0.8}Bi_2$ compound

Figure 2: Temperature variation of both real and imaginary *AC* magnetic susceptibility χ(*T*) for $CeNi_{0.8}Bi_2$ compound

Figure 1

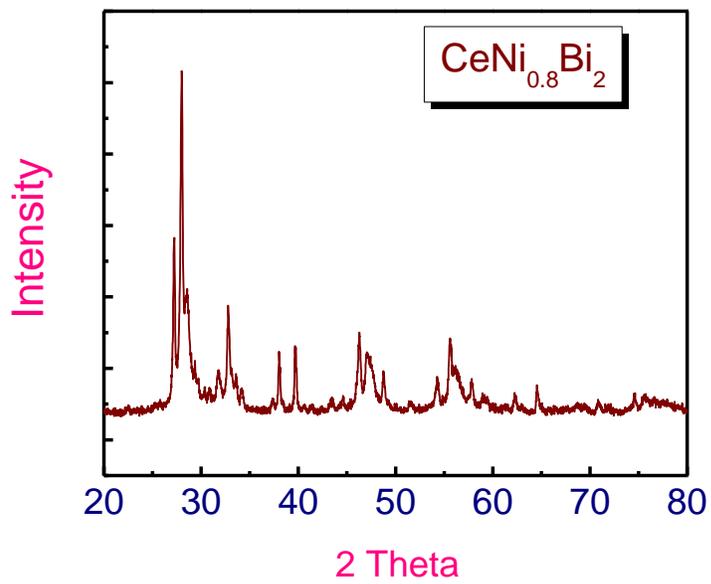

Figure 2

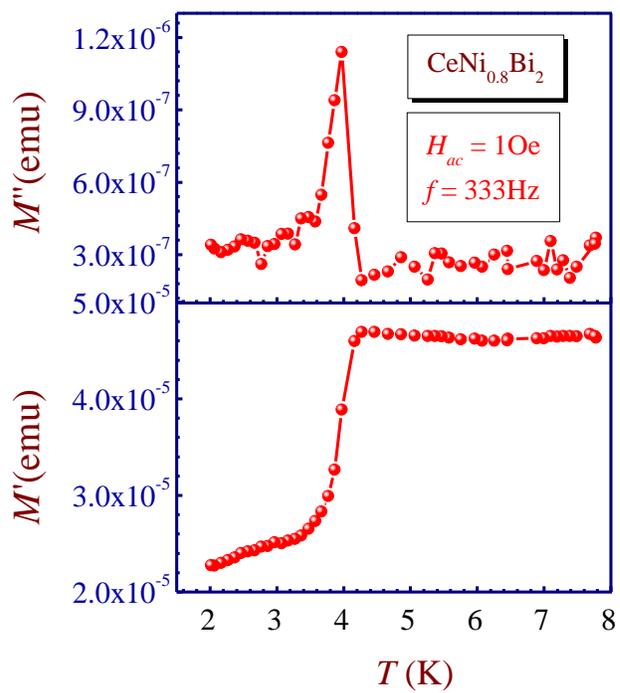